# On the importance of experimental details: A Comment on "Non-Polaritonic Effects in Cavity-Modified Photochemistry"

*Tal Schwartz*[1]   *and   James A. Hutchison*[2]

[1] School of Chemistry, Raymond and
Beverly Sackler Faculty of Exact
Sciences and Center for
Light–Matter Interaction,
Tel Aviv University
Tel Aviv 6997801 (Israel)
E-mail: talschwartz@tau.ac.il

[2] School of Chemistry and Australian
Research Council Centre of
Excellence in Exciton Science
The University of Melbourne
Masson Rd, Parkville, VIC, 3052
(Australia)
E-mail:
james.hutchison@unimelb.edu.au

**Recently, an article by the Barnes group reported on the experimental study of a photoisomerization reaction inside an optical cavity, claiming to reproduce previous results by Hutchison *et al.* and making the point that in such setups, changes in the absorption of ultraviolet radiation by the molecules in the cavity can lead to modifications in the photochemical reaction rate. While Hutchison *et al.* associated such modifications with the emergence of strong light-matter coupling, in their attempt to re-examine these experiments, Barnes *et al.* did not find any evidence that strong coupling needs to be invoked to explain the observed effects. In response to this publication, we herein highlight the main differences between the two experimental studies, and explain why the results of Barnes *et al.* are irrelevant to the former study and have no bearing on its conclusions. Specifically, we show that under the experimental conditions used by Hutchison *et al.* such intensity-modification effects are negligible and can therefore be ruled out.**

## 1 | Introduction

Just over a decade ago, we reported on the experimental study of a photoisomerization reaction taking place inside an optical cavity under strong coupling conditions[1]. Interestingly, in this work we found that under the appropriate conditions, the reaction rate observed for the molecular reaction within the cavity was slower than the rate of the reaction taking place outside of the cavity by up to a factor of four. Based on our experimental findings and data analysis, we concluded that the modification of the rate of the photochemical reaction resulted from the emergence of strong light-matter coupling, that is, the hybridization between the molecular wavefunctions and the photonic mode of the cavity, under resonant interaction between them. This work has sparked a surge of interest in strong coupling with molecules, and led to the emergence of "polaritonic chemistry", a rapidly expanding field in which strong light-matter coupling is employed to modify chemical reactions and materials' properties[2–8]. While polaritonic chemistry still presents several fundamental questions that need to be resolved[9,10], since this first experiment, chemical modifications under strong coupling have been observed in a plethora of molecular systems, both in photochemical[7], as well as in ground-state (i.e., thermally activated) reactions[11]. Specifically, changes induced by strong coupling in photoisomerization reactions have also been observed by other groups both in experiments[12,13] as well as in simulations[14], confirming that such reactions are indeed modified in the strong coupling regime. In a recent publication by Barnes and coworkers[15], the original photochromic system studied in reference 1 was revisited. The authors of reference 15 have claimed to reproduce our original results, however, based on their measurements and supplemental simulations, the authors have concluded that the changes in the rate of the photo-induced reaction can be fully explained by taking into account the influence of the cavity on the absorbance of light by the molecules embedded in it (that is, the photon absorption rate). While such trivial effects may, in general, take place in optical cavities, we stress that in reference 1 specific precautions were taken to avoid such artifacts, as provided in the Supporting Information section of reference 1 and as elaborated below. Furthermore, as we explain in this Comment, the sample preparation and experimental conditions of the two studies are vastly different, to the extent that the conclusions of reference 15 cannot be applied to interpret our previous results. In other words, the authors of reference 15 have conducted a different experiment, and, not surprisingly, have obtained different results.

## 2 | Polaritonic Effects in Cavity–Modified Photochemistry

Let us first summarize the main findings of reference 1. The photochemical system employed in this study was the spiropyran-merocyanine photochrome shown in Figure 1. Upon irradiation with ultraviolet (UV) light, the spiropyran (SPI) molecules, which are transparent over the entire visible range, undergo ring-opening and change their conformation to the merocyanine (MC) isomer which has a strong absorption band around 560 nm. This absorption band corresponds to an electronic transition, which was coupled with the cavity mode[1]. This photochromic reaction, as depicted in Figure 1a, is reversible, in the sense that photoexcitation of the MC molecules can induce backward-photoisomerization, transforming the molecules from the MC isomer back into SPI. This may be driven directly by absorption of 560 nm photons. However, it is also important to stress that **the MC isomer also absorbs in the UV** (as seen in Figure 1b), which, following internal conversion and relaxation to the lowest electronic excited state, may transform back to the SPI form. As detailed in reference 1, under continuous excitation and assuming **isosbestic irradiation conditions** (that is, at a wavelength for which the two isomers have the same absorption, which is around ~330 nm for this specific SPI/MC photochrome, see further discussion below), the kinetic equations for the photochromic system reduce to

$$\frac{d[\text{MC}]}{dt} = -(k_f + k_b)[\text{MC}] + k_f[\text{SPI}]_0 \quad (1)$$

where [MC] is the MC concentration at a given time, $k_f$ and $k_b$ are the effective forward and backward rate constants expressing the various microscopic molecular processes involved in the photoisomerization[1], and $[\text{SPI}]_0$ is the initial SPI concentration. The solution to this simple equation is $[MC](t) = [MC]_\infty(1 - \exp\{k_{obs}t\})$, with the observed reaction rate being $k_{obs} = k_f + k_b$ and $[MC]_\infty = [SPI]_0 k_f/(k_f + k_b)$ being the MC concentration at the photostationary state. Note that this solution also implies that at the photostationary state the SPI concentration is $[SPI]_\infty = [SPI]_0 k_b/(k_f + k_b)$ (neglecting irreversible photodegradation effects). Therefore, continuous UV irradiation normally leads to **a photostationary mixture of the two isomers**, with concentrations that are determined by the balance between the forward (SPI to MC) and backward (MC to SPI) photoisomerization processes, **both of which are driven by the same UV source**. This important detail is often overlooked, **which gives the (wrong) impression that under UV irradiation the reaction is unidirectional**, proceeding only from the SPI isomer to the MC one.

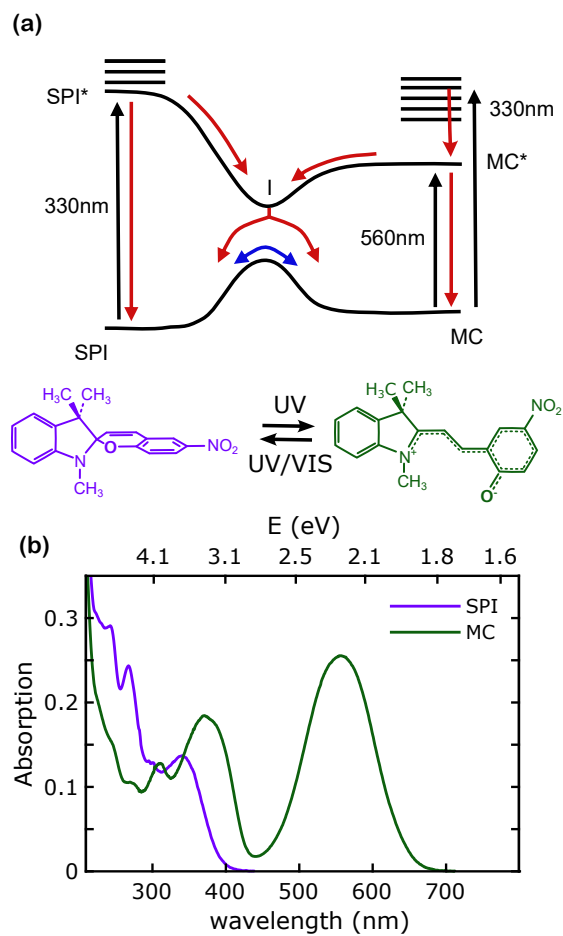

**Figure 1.** Photoisomerization of spiropyran molecules. (a) Schematic representation of the various pathways involved in the reversible photoisomerization process. UV light (330 nm used in reference 1) can excite the SPI isomer, which may either relax to its ground state or undergo photoisomerization via an intermediate state (I), ending at the MC ground state. The reverse process can be driven either by visible light excitation of the MC first excited state (~560 nm), or by UV excitation into higher excited states, followed by internal conversion into the lowest MC excited state. As indicated by the blue arrow, the reaction can also occur the electronic ground state, in which case it is thermally activated (with the SPI being the thermodynamically favoured isomer), however, under typical irradiation conditions the contribution of this pathway to the overall kinetics can be neglected. (b) Absorption spectra of a spin-coated film (with a PMMA host) in the SPI and MC forms.

The hybrid cavity, sketched in Figure 2a is comprised of a film of SPI molecules within a host polymer (polymethyl methacrylate, PMMA) in a 1:1 mass ratio, which was placed between two thin (~30 nm) silver layers serving as partially transmitting mirrors (see full details in reference 1). The top silver layer covered only half of the sample area, such that the other half was used as a non-cavity (i.e., bare molecules) reference. In addition, the photochromic layer was separated from the Ag mirrors by thin PVA buffer layers (20 nm), to avoid any chemical interaction between the molecules and the metal layers.

When the sample was irradiated with UV light (at 330 nm), the SPI molecules gradually converted into the MC form. The interaction strength between the cavity mode and the molecules is normally characterized by the (collective) Rabi splitting, which can be measured spectroscopically, and which scales as the square root of the concentration of molecules interacting with it (that is, the merocyanine molecules). Accordingly, the initial coupling strength was zero (when all the molecules were in their SPI form) and it increased as a function of UV irradiation time, as the molecules were gradually converted into the MC form, until reaching the photostationary state.

Figure 2b shows the measured transmission spectra in the initial (pure SPI, purple line) and final (photostationary, green line) states, when the cavity mode is tuned close to 560 nm, such that it is resonant with the MC transition to its first excited state. When the molecules are in the SPI state, the 'empty' cavity mode is seen as a single transmission peak. As the sample is irradiated and the molecules undergo photoisomerization to MC, the strong coupling between the MC molecules and the cavity mode results in the gradual splitting of this transmission peak[1,16]. In the photostationary state, after most of the molecules have been converted to MC, the transmission shows two polaritonic peaks, with an energy splitting of ~0.7 eV between them. In comparison, the reference (non-cavity) structure, whose spectra are shown in Figure 2c, exhibits an almost featureless transmission spectrum when the molecules are in their SPI state. This spectrum resembles that of thin Ag films in the visible range, slightly modified by the transparent dielectric layers located on top of the Ag mirror. Following irradiation, two broad dips appear in the transmission spectrum, around 560 nm and 380 nm. These dips correspond simply to the absorption peaks of the MC, reducing the transmitted intensity through the sample at these wavelengths. While it was previously suggested that strong coupling can take place even without the top mirror[17] (relying on reflections from the interface between the organic layer and air to form the cavity), **it is quite easy to see that here this is not the case**. Therefore, the non-cavity structure can be safely used as a reliable reference, provided that it is properly designed. For the off-resonance samples, the cavity (Figure 2d) is thick enough to support two modes. While the first-order mode is red-shifted with respect to the MC absorption (to $\lambda$ >900 nm), a second-order mode appears at 472 nm when the molecules are in the SPI form, and slightly shifts to 451 nm after the conversion to MC, due to the different refractive indices of the two isomers. Both of these modes are detuned from the MC absorption by more than the coupling strength ($g = 0.35 eV$), even when considering the angular dispersion of the first order mode (which shifts to 725 nm for an incident angle of 90°, as obtained by transfer matrix calculations). For the thick reference sample (Figure 2e), when the molecules are the in SPI form one can indeed observe a broad transmission peak around 480 nm (0.38 eV above the MC absorption), which can be associated with the 'cavity-free mode' discussed in reference 17. However, the width of this 'mode' is roughly 0.8 eV (corresponding to a Q-factor of ~3), and therefore it does not fulfill the conditions for strong coupling. As such, the two transmission maxima appearing after the conversion to MC should be regarded more as the result of the MC absorption cutting a dip in the transmission of the cavity-free mode (in a similar manner to Figure 2c) rather than polaritonic peaks. This conclusion is also supported by the calculated absorbance (see Figure 5p), which indicates that this system is still below the threshold for strong coupling. We note that samples which are much thicker may display higher Q-factors, and may exhibit cavity-free strong coupling, under similar conditions.

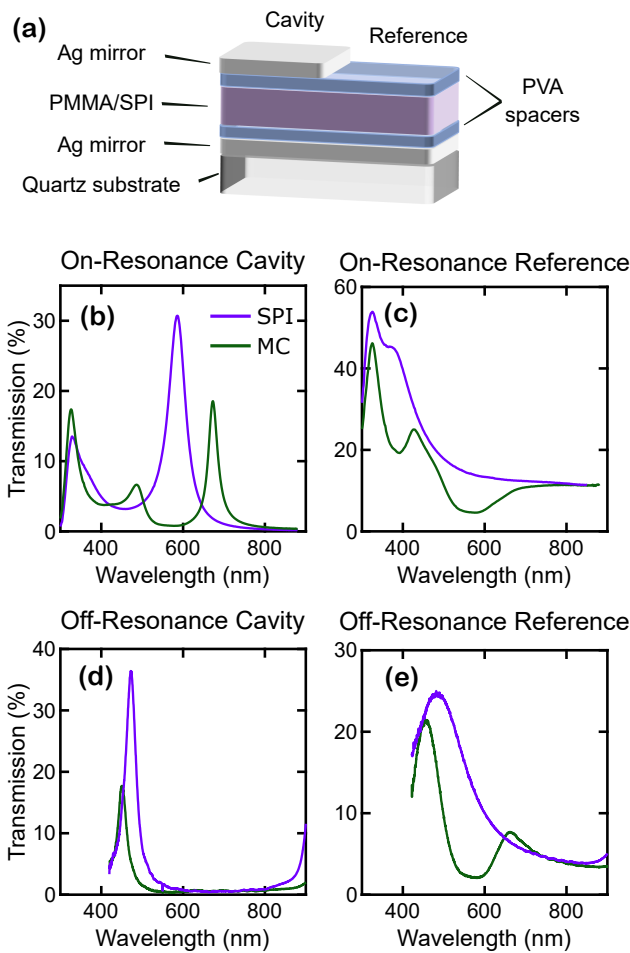

**Figure 2.** (a) Structure of the sample used in reference 1. The cavity side is composed of a PMMA/SPI film between two Ag mirrors (and separated from the metal mirrors by ~20 nm PVA layers). The reference side is identical to the cavity side, apart from the missing top Ag mirror, which is present only in the cavity side. (b)-(e) Transmission spectra measured for on-resonance cavity (b) and reference (c) or off-resonance cavity (d) and reference (e), in the SPI and MC forms. The total thicknesses of the organic layers are ~130 nm in the on-resonance sample and 220 nm in the off-resonance sample.

By monitoring the evolution of the transmission spectrum of the hybrid cavity during the irradiation and comparing to transfer-matrix calculations, we extracted the MC concentration as a function of time to obtain the photoisomerization kinetics. Then, by plotting the concentration as a function of time on a logarithmic scale (see Figure 3), the reaction rate was estimated from the slope of the curve. When comparing the results of this analysis for the resonant cavity and the non-cavity structure, we found that the reaction rate within the cavity (red dots) was significantly lower than the rate observed in the reference (blue dots).

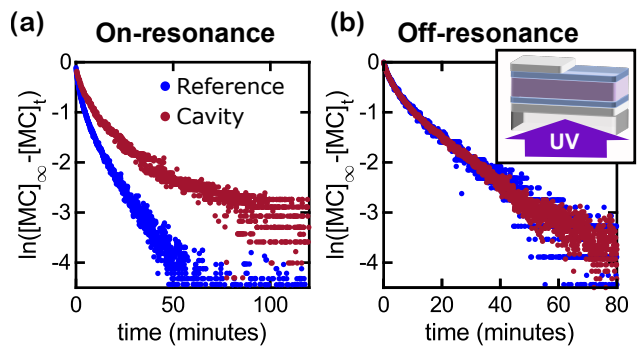

**Figure 3.** Photoisomerization kinetics measured inside the cavity (red) and on the reference side (blue) of the tuned (a) and red–detuned (b) samples, showing the slow–down of the photoisomerization under resonant strong–coupling conditions. The rate of the reaction can be deduced assuming a first–order reaction whose kinetics are described by $[MC](t) = [MC]_\infty (1 - \exp\{k_{obs}t\})$, where $[MC](t)$ is the MC concentration at time $t$, $[MC]_\infty$ the concentration at the photostationary state and $k_{obs}$ is the observed rate of the photochemical reactions (which may vary with time). It is important to note that the measurements were carried out with the UV irradiation impinging from the substrate side, as illustrated in the inset, such as to balance the irradiation intensities reaching the photochromic layer on either the cavity or reference side.

Approaching the photostationary state, when the MC concentration (and hence the coupling strength) reaches its maximal value, the slope of the curve describing the kinetics in the cavity is roughly four-fold smaller than the slope corresponding to the reference. Furthermore, for the off-resonance structures the difference between the kinetics of the cavity and non-cavity samples vanished. These results led to the conclusion that strong coupling between the cavity and the MC molecules modified the dynamics of the photochemical reaction.

As explained above, the observed rate of such a bi-directional reaction (as extracted from the kinetics in Figure 3) represents the **rate at which the system reaches the photostationary state**, which involves **the rates of both the forward and backward processes, driven by the same UV source and occurring simultaneously**. Since it is the MC excited-state that is coupled with the cavity mode, it is reasonable to assume that **only the backward photochemical process is affected by strong coupling**. Nevertheless, by examining the kinetics of system, as expressed by Equation 1 it is easy to see that **any change in the microscopic rate-constant of the photoinduced backward reaction will also be expressed in the observed rate**, as was found experimentally. We stress that, as illustrated by the inset of Figure 3, the samples were irradiated through the substrate (bottom) side, in order to balance the UV intensities experienced by the photochromic molecules in both types of structures (see extended discussion below). It is also worth mentioning that in the on-resonance cavity, the influence of the cavity on the reaction rate (that is, the difference between the cavity and non-cavity kinetics) was seen to increase gradually, consistent with the fact that the coupling strength increases as a function of irradiation time.

While these measurements and the experiments conducted in reference 15 appear similar at first sight, there are several important differences which make the two studies significantly distinct from each other, as explained below. For that reason, the conclusions of Barnes and coauthors do not apply to the results of our original study.

## 3 | Discussion

When studying the kinetics of photochemical reactions, care must be taken to avoid spurious effects on the reaction. This is even more important when one tries to isolate the effect of strong coupling on the reaction. For that reason, the samples and the experiments in reference 1 were carefully designed and performed under specific conditions. For example, the top Ag mirror of the cavity was not deposited directly on the molecules, since the deposition process itself could affect the chemical properties of the molecules or the host polymer. Instead, we employed a stamping method for adding the top mirror and closing the cavity, as described in reference 1. Of further importance, **specific conditions were chosen for the optical design and the irradiation of the samples**. In reference 1, we employed a combination of Ag mirrors with irradiation at 330 nm for inducing the photochemical reaction (using an Hg source and a bandpass filter). While Ag is highly reflective in the visible range, its reflectivity drops off sharply below ~400 nm. In particular, using 330 nm irradiation has the advantage that it overlaps with a transparency window associated with the bulk plasmon of Ag at 326 nm[18], with a transmission of ~60% and reflection of 10-20% at that spectral region. As a result, these structures could support any cavity mode in the UV region, which could lead to an enhanced absorption rate. Moreover, as pointed above, this wavelength is close to an isosbestic point of the photochromic molecules. The lack of an optical resonance and the similar absorbance of the molecules within the cavity resulted in an optical response (for the 330 nm irradiation) which remained almost invariant throughout the measurement and **a near-constant flux of photons inside the optical cavity, irrespective of the extent of photoisomerization**. This can be seen by examining the transmission spectra around this wavelength (Figure 2a), and also by a more detailed analysis, as discussed below. Employing isosbestic irradiation is common practice in studying the kinetics of photochemical reactions. In particular, in this class of photochromic reactions represented by the SPI/MC system, the simultaneous occurrence of the forward and backward processes has serious consequences for resolving a unique kinetic solution, and the simple dynamics described by Equation 1 are only valid under monochromatic excitation at an isosbestic point[19,20]. Irradiation away from an isosbestic point means that the penetration depth into the sample changes as the photochemical reaction proceeds, resulting in varying photon-flux and varying excitation rates of the photochrome. This can lead to complex kinetics[20,21], which must be further complicated when the photochrome is then studied inside an optical cavity, making meaningful comparison risky. Broadband illumination, as conducted in Reference 17, drastically complicates this issue, as the (unresolved) molar absorptivity of MC and its photochromic quantum yield vary with wavelength.

Apart from the sample design, by irradiating the sample from the bottom side, as illustrated in the inset of Figure 3, we could ensure that **the photon fluxes reaching the molecular layer on both the cavity side and the reference side were almost identical**, since in both cases the UV light passed through the same Ag layer. Keeping in mind the low reflectivity of the Ag mirrors at 330 nm, one may estimate the difference in the photon flux densities between the cavity and non-cavity (reference) sides. In the non-cavity case, the UV light, impinged on the sample through the substrate (see right side of inset in Figure 3), passed through the bottom Ag mirror and then made a single pass through the molecular layer. On the other hand, on the cavity side (left side of the inset), the UV irradiation passed through the same bottom Ag mirror, but was then partially reflected (~20%) from the top mirror back into the molecular layer. With this low reflectivity, further reflections of the light can be safely neglected, and hence the effect of the cavity on the UV irradiation amounted to a relatively small increase in intensity (and therefore increased energy absorption rate) of about 20%. To show this explicitly, we provide in Figure 4 the energy-dissipation distribution (see reference 16) for various conditions, calculated using transfer matrix simulations (further details are provided in the Appendix and in Figure 5). As can be seen in Figure 4a, when the samples are irradiated from the substrate side (and through the Ag mirror), the absorption rate across the cavity (solid lines) and reference (dashed lines) structures are quite similar to each other. This occurs both when the molecules are in the SPI form (purple lines) as well as in their MC form (green lines). Although the two types of structures exhibit different distributions of the absorbed power (resulting from interference between the incoming field and the weak field reflected from the top Ag layer), when integrated over the entire height of the photochromic layer, the difference between the structures amounts to the ~20% increase in the cavity, as mentioned above. This is also the case for the off-resonance samples, as can be seen in Figure 4b. In sharp contrast, when the UV light is injected from the air side (Figure 4c), the energy-dissipation is quite different between the cavity and reference structures, with approximately double the integrated absorption rate in the reference (non-cavity) structure (see dashed lines *vs* solid lines). This, of course, would prevent any meaningful comparison between the cavity and the reference, which is why we chose to irradiate the samples through the substrate. **This was not the case in the recent reference** 15.

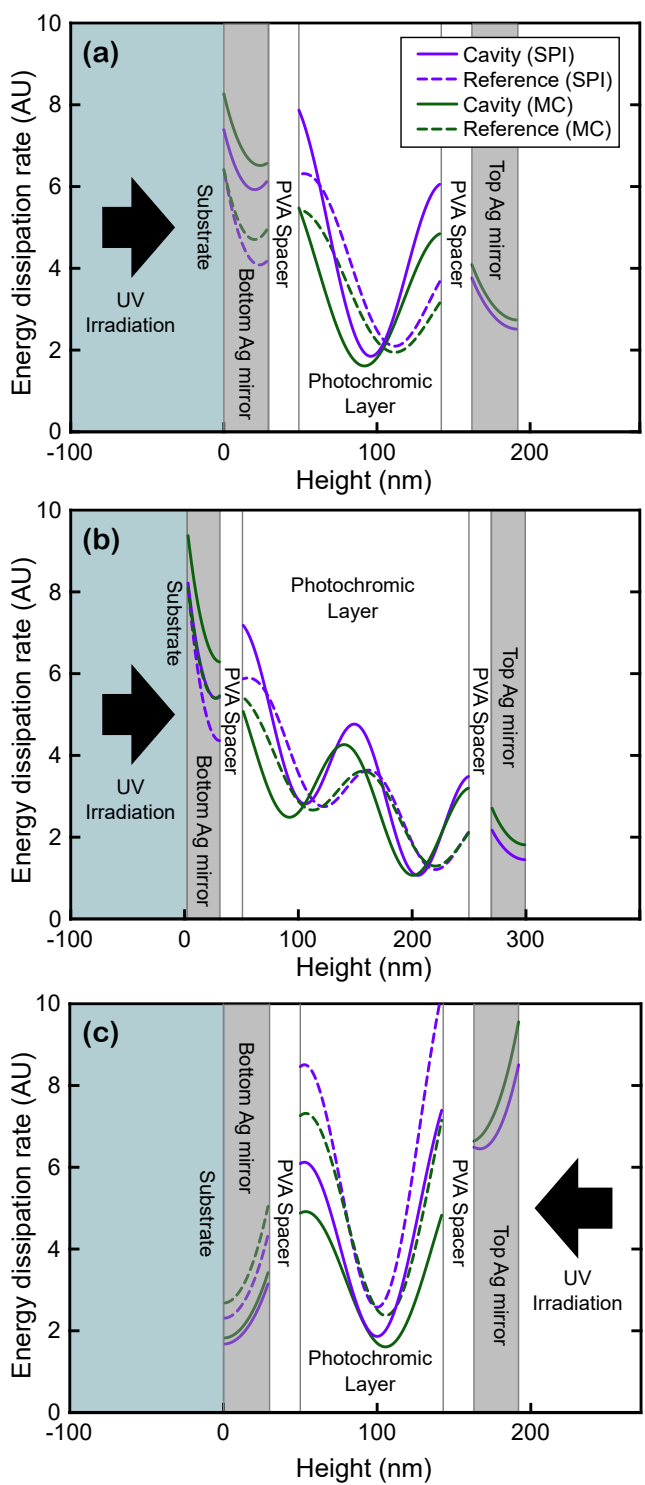

**Figure 4.** (a) Distribution of UV energy dissipation within the cavity (solid lines) and reference (dashed lines) structures, obtained from transfer matrix simulations and with the irradiation coming through the substrate (note that the top mirror is absent in the reference structure). As seen, in the SPI and MC forms, the UV absorption rates display somewhat different distributions, but with overall similar magnitudes. The data presented here is based on the results shown in the Appendix (Figure 5) by integrating the spectral density of absorption rate over the range of 325–335nm. (b) Same as (a) but calculated for the red-detuned structure. (c) Same as (a) but with the structures irradiated from the air side. Here the absorption rate in the reference structure is roughly twice as high as that in the cavity.

The main point conveyed by Barnes *et al.* was that, since the rate of any photochemical process depends on the photon-absorption rate, any modification in local intensities will lead to a change in the photoisomerization rate[15]. Therefore, working under the irradiation conditions described above, as well as using Ag mirrors, is crucial for avoiding such trivial effects.

In reference 15, the authors mainly focused on measurements in cavities composed of Al mirrors, which have high reflectivity in the UV region. This can indeed give rise to a resonant enhancement of the UV light driving the reaction and make irradiation at the isosbestic point difficult, which is obviously undesired. Moreover, the irradiation was done using a broadband Xe arc lamp, which covered the entire range from the UV to the near infrared (see Figure S3 in the Supporting Information section of reference 15). The complication this causes for meaningful analysis of SPI/MC photochromism outside the cavity has been discussed above. To repeat, under such conditions, absorption of visible light will cause a direct excitation of MC molecules or, once strong coupling is reached, direct excitation of polaritons. This, in turn, can activate another channel for the backward photoisomerization from MC to SPI, which may further complicate the kinetics. Even more, since the cavity transmission in the vicinity of 560 nm depends on the concentration of MC molecules (through the gradually increasing coupling strength and spectral splitting), the photon flux driving this photochemical pathway, and therefore the rate of this process most probably varies during the course of the measurements.

Finally, during the UV irradiation, both excited isomers may experience irreversible photo-oxidation processes through triplet formation[22], which leads to degradation and competes with the light-induced reversible switching between the two isomers. To avoid such spurious effects, in reference 1 we conducted the kinetic photochemical measurements with the sample held inside a vacuum chamber, at a pressure of $10^{-3}$ mbar. In contrast, the studies in reference 15 were performed under normal atmospheric conditions, where the presence of such photo-oxidation pathways can significantly alter and complicate the kinetics. While such technical details may seem minor, we stress that the experimental conditions and the specific details of the samples used in our experiments were specifically chosen following careful optimization accompanied by numerical analysis. It is therefore not surprising that deviating from these conditions, as in reference 15 would give rise to various side-effects.

## 4 | Conclusions

In summary, while Barnes *et al.* have conducted extensive measurements involving a multitude of cavities and several types of mirrors, their findings regarding the potential influence of the cavity on the intensity of light driving the reaction is quite anticipated. In fact, this idea is similar in spirit to old concepts such as cavity-enhanced spectroscopy[23] or plasmon-enhanced photochemistry[24,25], and can be regarded as a weak coupling effect. Since such intensity-enhancement effects can indeed obscure the effect of strong coupling on the reaction, in our original work we chose very specific conditions so as to minimize the impact of the optical structure on the UV irradiation intensity. The first step in achieving such conditions is to avoid strong reflectivity in the UV region, for example by employing Ag mirrors, as in our work. As we pointed out in reference 1, and as we have shown here explicitly, the weak reflectivity of the Ag mirrors does lead to a small enhancement by the cavity at 330 nm, amounting to ~20% higher intensity relative to the reference structure. While such enhancement can indeed lead to a small acceleration of the photochemical reaction, it cannot explain the slowing down observed in our experiments, which reached **a four-fold reduction** in the photochemical reaction rate.

As the effects of strong coupling on chemical reactions are still far from being understood, polaritonic chemistry calls for extensive and systematic experimental investigation, which will shed more light on the mechanisms at play. However, this must be done under rigorous conditions, such as in references 1,12,13 where the strong coupling effects on photoisomerization are clearly visible. While reference 15 seems to present a systematic study, the authors did not take the elementary precautions described in our original study, resulting in artifacts which completely overwhelmed the effect of strong coupling on the reaction.

## APPENDIX

To understand the effect of the optical structure on the photon absorption rate in the photochrome layer, we used transfer-matrix simulations and calculated the spectral response under the various conditions. As described in reference 16, we first calculated the field distribution $|E(z)|$ within the structure as a function of photon energy. As shown in Figure 5, we calculated the field distributions (normalized with respect to the incoming field strength and with the light coming from the top side) for the full cavity with the molecules in both SPI and MC forms, and repeated these calculations for the reference structures having a single Ag mirror. For the on-resonance cavity and with the molecules in the SPI form (Figure 5a), one can clearly observe the field-enhancement effect around 2.2 eV, corresponding to the ('empty') cavity mode. In the MC cavity (Figure 5b), this mode splits into the two polaritonic states, which show up as two lobes around 1.84 and 2.54 eV. In contrast, in the UV (> 3.1 eV) the field does not exhibit any resonant enhancement in any of the structures, which is consistent with our expectation that the Ag mirrors cannot support a cavity mode in the UV spectral region. Finally, in the reference structures (Figure 5c,d), no resonant mode is observed through the entire spectral range, which confirms that the single-mirror structure can indeed function as a non-cavity reference where strong coupling effects cannot appear. By multiplying the local intensity ($|E|^2$) by the imaginary part of the dielectric function, we obtain the distribution of the energy dissipation, as a function of position and photon energy, as shown in the second row of Figure 5 (panels e-h). Then, by spatially integrating this distribution over the region containing the photochromic molecules only, we obtain the probability for photon absorption by the molecules, which is depicted by the red curves in Figure 5e-h (here a value of one corresponds to complete absorption). As seen, in the UV region, and in particular at 3.75 eV (330 nm), roughly 25% of the incoming photons are absorbed by the molecules inside the cavity (both in the SPI and MC form), whereas in the non-cavity structures the absorption probability reduces to ~20%. From this analysis we conclude that the cavity enhances the energy absorption rate by about 20%, as pointed out earlier.

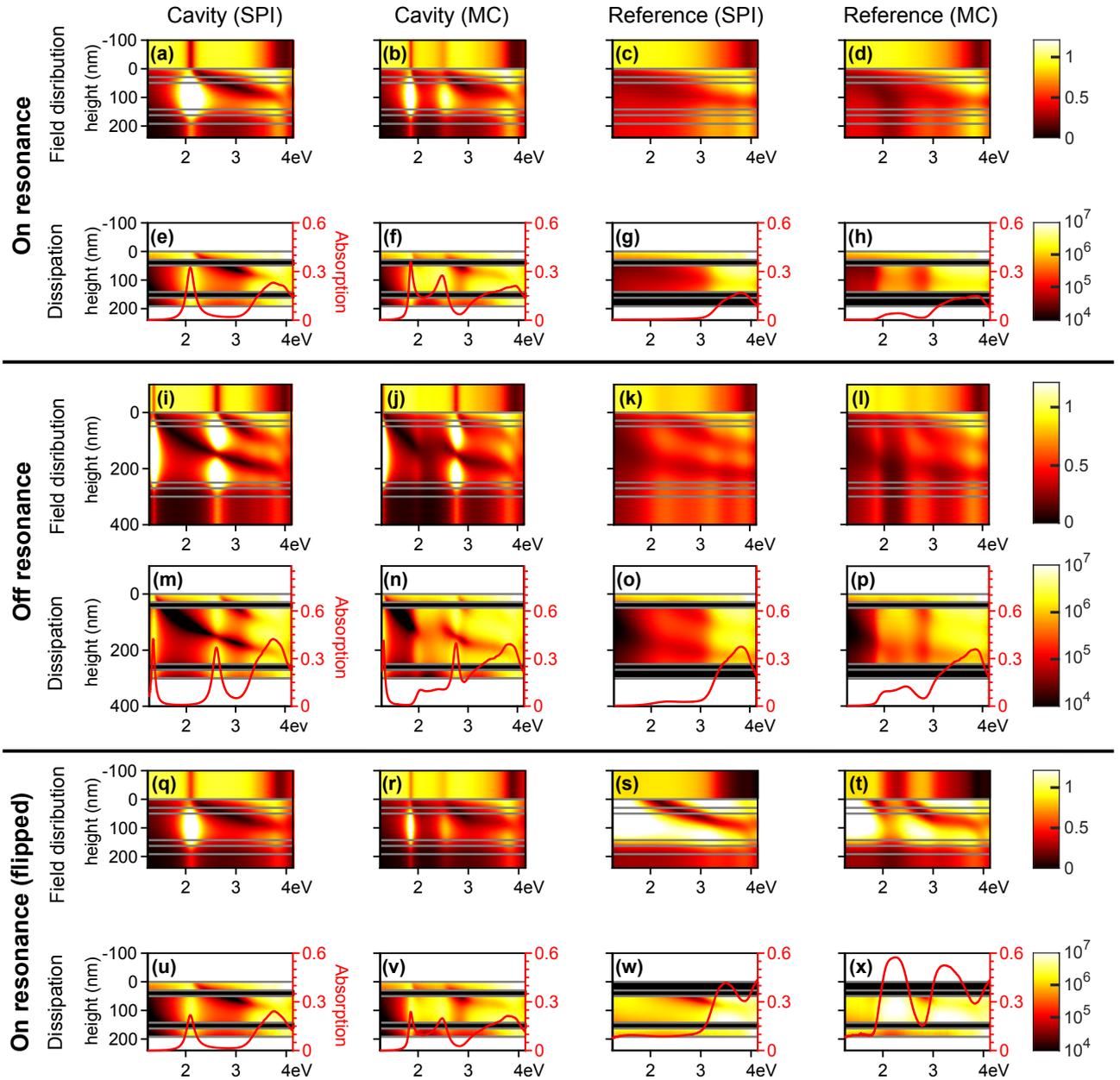

**Figure 5.** Spectral analysis obtained by transfer-matrix simulations, for the on-resonance sample (a–h) and off-resonance sample (i–p), both irradiated from the substrate side and the on-resonance sample irradiated from the air side (q–x), calculated with the order of the layer reversed. For the on-resonance case (a–h), the top panels show the field strength distribution $|E|(z,\omega)$ across the structure (with the horizontal gray lines marking the edges of the various material layers) as a function of photon energy, for an incoming intensity of 1. Note that the incoming wave was removed from the image, such that the distribution on the top side of the structure shows only the reflected field. The bottom panels show the distribution of energy dissipation rate, which is given by $P_d = \omega \epsilon''(\omega)|E(z,\omega)|^2$, with $\omega$ being the light angular frequency and $\epsilon''$ being the imaginary part of the (frequency-dependent) dielectric function at each position. The red lines represent the total power (as a function of photon energy) absorbed by the photochromic molecules, obtained by integrating $P_d$ over the entire thickness of the molecular layer. As in the experiment, here the light impinges the sample from the substrate side. Importantly, for irradiation at 3.75 eV (330 nm), the absorbed power is similar under all four types of conditions (cavity/reference in SPI or MC form) with the cavity displaying somewhat higher (by ~20%) absorption compared to the reference. (i–p) Same as (a–h), but calculated for the off-resonance (red-detuned) sample. (q–x) Same as (a–h), calculated for the on-resonance sample but with the light impinging on the sample from the air side. Note that under such conditions the absorption (at 3.75 eV) in the cavity and in the reference are considerably different.

Repeating this analysis for the off-resonance structures (Figure 5i-p) reveals similar behavior. The cavity modes (now detuned from 2.2 eV) give rise to field-enhancement in the visible part of the spectrum only (see Figure 5i,j), which does not appear in the reference structures (Figure 5k,l). Moreover, inspecting the absorption probability (Figure 5m-p, integrated over the photochromic layer only) we find that here the differences between the various samples at 3.75 eV are even smaller (albeit being somewhat higher than in the on-resonance case). This again confirms that comparing the reaction rates of the cavity and non-cavity structures provides a reliable means to isolate the effect of strong coupling on the reaction quantum yield.

Finally, we perform the same analysis on the resonant structures, but consider the case in which the light enters the samples from the air side, in contrast to excitation from the substrate side (Figure 5q-x). Here, although the cavity gives rise to similar field-enhancement at the resonant modes (see Figure 5q,r), the absorption probability across the whole spectrum is much higher in the reference structures than in the cavity structures, as seen in Figure 5u-x. This demonstrates that, in addition to the structure design and the excitation wavelength, the direction of excitation is also important for balancing the excitation rates between the cavity and the reference samples, as discussed above.